\documentclass[11pt]{article}
\usepackage{graphicx}
\usepackage{amssymb}
\usepackage{amscd}
\usepackage{mathrsfs}
\usepackage{longtable,lscape}
\usepackage{amsthm}
\usepackage{amsfonts}
\usepackage{amsmath}
\usepackage{bbm}
\usepackage{float}
\usepackage{url}

\newcommand{\captionfonts}{\footnotesize}
\makeatletter  % Allow the use of @ in command names
\long\def\@makecaption#1#2{%
  \vskip\abovecaptionskip
  \sbox\@tempboxa{{\captionfonts #1: #2}}%
  \ifdim \wd\@tempboxa >\hsize
    {\captionfonts #1: #2\par}
  \else
    \hbox to\hsize{\hfil\box\@tempboxa\hfil}%
  \fi
  \vskip\belowcaptionskip}
\makeatother 
\begin{document}
\title{\bf Modeling Human Decision-making: An Overview of the Brussels Quantum Approach} 
%\author{Diederik Aerts$^1$, Massimiliano Sassoli de Bianchi$^{1}$, Sandro Sozzo$^{2}$  and Tomas Veloz$^1$ \vspace{0.5 cm} \\ \normalsize\itshape $^1$ Center Leo Apostel for Interdisciplinary Studies, Brussels Free University \\  \normalsize\itshape Krijgskundestraat 33, 1160 Brussels, Belgium \\ \normalsize\itshape $^2$ School of Business and Research Centre IQSCS, University of Leicester \\  \normalsize\itshape University Road, LE1 7RH Leicester, United Kingdom \\}
\author{Diederik Aerts$^1$, Massimiliano Sassoli de Bianchi$^3$, Sandro Sozzo$^3$ and Tomas Veloz$^4$\vspace{0.5 cm} \\ 
\normalsize\itshape
$^1$ Center Leo Apostel for Interdisciplinary Studies and Department of Mathematics\\
\normalsize\itshape 
Brussels Free University, Krijgskundestraat 33, 1160 Brussels (Belgium)\\
\normalsize
E-Mail:  \url{diraerts@vub.ac.be}
\vspace{0.3 cm} \\ 
\normalsize\itshape
$^2$ Center Leo Apostel for Interdisciplinary Studies \\
\normalsize\itshape 
Brussels Free University, Krijgskundestraat 33, 1160 Brussels (Belgium) and\\
\normalsize\itshape
Laboratorio di Autoricerca di Base, via Cadepiano 18, 6917 Barbengo (Switzerland)\\
E-Mail:  \url{msassoli@vub.ac.be}
\vspace{0.3 cm} \\ 
\normalsize\itshape
$^3$ School of Business and Centre IQSCS, University of Leicester \\ 
\normalsize\itshape
University Road, LE1 7RH Leicester (United Kingdom) \\
 \normalsize
        E-Mail: \url{ss831@le.ac.uk} 
        \vspace{0.3 cm} \\ 
\normalsize\itshape
$^4$
Universidad Andres Bello, Departamento Ciencias Biol\'ogicas\\ 
\normalsize\itshape
Facultad Ciencias de la Vida, 8370146 Santiago (Chile) and \\
\normalsize\itshape
Instituto de Filosof\'ia y Ciencias de la Complejidad\\
\normalsize\itshape
Los Alerces 3024, \~Nu\~noa, Santiago (Chile)\\
\normalsize
        E-Mail: \url{tveloz@gmail.com}
}
\date{}
\maketitle
\begin{abstract}
\noindent 
We present the fundamentals of the quantum theoretical approach we have  developed in the last decade to model cognitive phenomena that resisted modeling by means of classical logical and probabilistic structures, like Boolean, Kolmogorovian and, more generally, set theoretical structures. We firstly sketch the operational-realistic foundations of conceptual entities, i.e. concepts, conceptual combinations, propositions, decision-making entities, etc. Then, we briefly illustrate the application of the quantum  formalism in Hilbert space to represent combinations of natural concepts, discussing its success in modeling a wide range of empirical data on concepts and their conjunction, disjunction and negation. Next, we naturally extend the quantum theoretical approach to model some long-standing `fallacies of human reasoning', namely, the `conjunction fallacy' and the `disjunction effect'. Finally, we put forward an explanatory hypothesis according to which human reasoning is a defined superposition of `emergent reasoning' and `logical reasoning', where the former generally prevails over the latter. The quantum theoretical approach explains human fallacies as the consequence of genuine quantum structures in human reasoning, i.e. `contextuality', `emergence', `entanglement', `interference' and `superposition'. As such, it is alternative to the Kahneman-Tversky research programme, which instead aims to explain human fallacies in terms of `individual heuristics and biases'.
\end{abstract}

\section{The quantum cognition research programme\label{intro}}
Researchers in cognitive and social science have assumed for many years, often implicitly, that human judgment and decision-making under uncertainty can be modeled by means of set theoretical structures, i.e. sets and operations between sets. These structures are very similar to those used in distributive logic, formalized by George Boole (`Boolean logic'), and probability theory, axiomatized by Kolmogorov (`Kolmogorovian probability') \cite{k1933}. There is agreement, in both the psychology and physics communities, to refer to these structures as `classical structures', as they were originally used in classical physics and later extended to psychology, computer science, statistics, economics, finance, etc. In decision theory, classical structures have been, again implicitly, incorporated into the so-called `Bayesian paradigm': any source of uncertainty can be formalized probabilistically in a Kolmogorovian sense. Expected utility theory (EUT), which provides the normative foundations of rational behavior under uncertainty, then guarantees that decision-makers should behave as if they maximized EU with respect to a Kolmogorovian probability measure formalizing their subjective uncertainty about the world \cite{s1954}.

More recently, this classical paradigm has been seriously challenged by a number of paradoxical findings in cognitive psychology, revealing that classical structures are generally unable to model concrete human decisions. This has made highly problematical the interpretation of a whole set of cognitive phenomena in terms of Boolean logic and Kolmogorovian probability. In regard to that, empirical  literature has identified the following two major types of `deviations from classicality'.

(i) `Probability judgment errors': people judge the probability of the conjunction `$A$ and $B$' (disjunction `$A$ or $B$') of two events $A$ and $B$ as higher (lower) than the probability of one of them, violating in this way the `monotonicity law of Kolmogorovian probability'.

(ii) `Decision-making errors': people prefer action $A$ over action $B$ if they know that an event $Z$ occurs, and also if they know that event $Z$ does not occur, but they prefer $B$ over $A$ if they do not know whether $Z$ occurs or not, violating in this way the `law of total Kolmogorovian probability'.

Errors of type (i) manifest, for example, in terms of `over/under-extension effects' in typicality and membership judgments on conceptual categorization \cite{os1981,h1988a,h1988b,ap2011}. But, the most famous evidence of probability judgment error was discovered by Kahneman (Nobel prize in economics 2002) and Tversky (one of the most cited psychologists) by means of their `Linda story' and is known as the `conjunction fallacy'  \cite{tk1983}.  The conjunction fallacy has then been systematically confirmed and  found  generating further effects, fallacies, paradoxes and contradictions (see, e.g., \cite{bpft2011,bb2012} for a review of cognitive fallacies). 

In 1992, Tversky and Shafir detected a significant example of decision-making error in the `disjunction effect' \cite{ts1992}, which led to the conclusion that people do not generally apply rational reasoning when taking concrete decisions but, in certain situations, they are instead strongly influenced by contextual aspects of a cognitive nature, like `uncertainty aversion'. The disjunction effect violates the law of total Kolmogorovian probability, and also the `sure thing principle', one of the fundamental axioms of EUT. A similar violation was observed in 1961 by Daniel Ellsberg, who noticed that people tend to avoid maximizing EU if they have to take decisions under an uncertainty aversion scenario (`Ellsberg paradox') \cite{e1961}.

A quite influential solution of the above difficulties came from Kahneman and Tversky themselves, who put forward that the above deviations are `genuine fallacies of human reasoning', elaborating a theory based on `judgment heuristics and individual biases' to explain them \cite{tk1974}. The Kahneman-Tversky solution, though valid at an intuitive level, does not seem to provide sufficient clarifications with respect to the deep mechanisms underlying human reasoning, hence it is generally accepted that efforts should be made to move beyond the fragmentation of existing approaches to develop a more coherent, comprehensive and deductive theoretical hypothesis \cite{so2008}.

An important alternative considers the traditional logical and probabilistic modelling schemes as too restrictive to capture the way in which the human mind works in a concrete judgment or decision. This approach therefore adopts, with great success, the more general mathematical formalism of quantum probability, which is known to be non-Kolmogorovian in a very precise sense. Impressive results have already been obtained by this `quantum cognition research programme', in the modelling of conjunction and disjunction fallacies, human similarity judgments, disjunction effects and Ellsberg-type paradoxes, knowledge and meaning representation, question order effects, etc. \cite{bpft2011,bb2012,a2009,abgs2013,ags2013,hk2013,pb2013,pnas,bwb2015,m2015,hk2016,kpsb2016,ahs2017,pbsy2017}.

We illustrate here the Brussels contribution to this novel, exciting and potentially cutting-edge research programme. In Section \ref{brussels}, we present the foundations of the operational-realistic approach to human cognition we have recently developed relying on a two-decade research on the axiomatic approaches to quantum physics, the origins of quantum probability and the appearance of quantum structures outside the micro-world. We show that any conceptual entity, that is, a single concept, a conceptual combination, a proposition, or a more complex decision-making entity, can be abstractly described in terms of the operationally defined notions of `states', `contexts', `properties', `measurements' and `outcome probabilities', and that impressive analogies recur between the operational description of measurement processes and the effects of context on quantum and conceptual entities. This indicates that the quantum formalism and its quantum probability structure are natural candidates to model cognitive phenomena. Then, we formulate in Section \ref{concepts} an explicit quantum theoretic approach to model combinations of two concepts, overcoming the classical modeling difficulties that arise in over- and under-extension effects on conceptual categorization. Next, we apply in Sections \ref{linda} and \ref{disjunction} the quantum theoretical approach to model the conjunction fallacy and the disjunction effect, respectively, explaining the empirical deviations from classical modeling in terms of genuine quantum structures. Finally, we formulate in Section \ref{explanation} a theoretical hypothesis which, on the one side explains emergence of quantum structures in cognition and, one the other side, provides a unifying and coherent explanation across the cognitive effects mentioned above. This explanatory hypothesis interprets human reasoning as a superposition of `logical reasoning' and `emergent reasoning', and research on cognitive fallacies  allows to demonstrate that the former generally prevails over the latter.

The Brussels quantum modeling approach to conceptual entities was the starting point for the development of a quantum theoretical approach to `meaning entities' that can be associated with 
 documents and texts, like those of the World Wide Web, which is presented in \cite{asdbsv2018}. %and constitutes a substantial part of the Brussels contribution to the Quantum Information Access and Retrieval Theory (QUARTZ) Consortium.\footnote{We refer to the webpage \url{http://www.quartz-itn.eu/about} for more information on the QUARTZ Consortium.}

\section{The Brussels operational-realistic approach to cognition\label{brussels}}
It is important to note that the modeling effectiveness of quantum theory regarding the situations mentioned  
in Section \ref{intro} is not necessarily due to the existence of microscopic quantum processes in the human brain, at least at the present stage of research. 

So, why should cognition go into quantum?

From our point of view, the reason for such a bold change of methodological perspective can be traced back to the studies on the axiomatic foundations of quantum physics, the origins of quantum probability and the differences between classical and quantum theories. More precisely, we 
were mainly guided by: 

(i) the deep analogies between quantum particles and conceptual entities with respect to `potentiality' and `contextuality';

(ii)  the role played by quantum probability in formalizing experimental situations where potentiality and contextuality are present.

It is indeed well known in quantum physics that, in a quantum measurement process, the measurement context physically influences the quantum particle that is measured in a `non-deterministic way', actualizing one outcome within a set of possible measurement outcomes, as a consequence of the interaction between the quantum particle itself and the measurement context. Suppose now that a statistics of measurement outcomes is collected after a sequence of many repeated measurement processes on  an arbitrary quantum entity, and such that (i) the measurement actualises properties of the entity that were not actual before the measurement started; 
(ii) different outcomes and actualisations are obtained probabilistically. What type of probability can formalize such experimental situation? It cannot be a Kolmogorovian probability,  because Kolmogorovian probability formalizes a situation of 
lack of knowledge about properties of the entity that were already actual before the measurement started. On the other hand, one 
can demonstrate that a situation where a measurement context actualizes properties that were only potential before the measurement started can be represented in a generalized quantum probabilistic framework \cite{a1986,p1989}.

How about a human decision process? We realized that a decision process is generally made in a state of genuine potentiality, which is not of the type of a lack of knowledge of an actuality. Consider, for example, a survey including the question ``are you a smoker or not?'', and suppose that 21 participants over a whole sample of 100 participants answer  `yes' to this question \cite{aa1995}. In the large number limit, $0.21$ is interpreted as the probability of finding a smoker in this sample of participants. This probability clearly formalizes a `lack of knowledge about an actuality', because each participant `is' a smoker or `is not' a smoker before the property 
was tested, hence before the experiment to test it -- the survey -- started. Consider now the question ``are you for or against the use of nuclear energy?'', and suppose that 31 participants answer `yes' to this question. In this case, the resulting probability $0.31$ cannot formalize a lack of knowledge about an actuality. Indeed, due to the nature of the question, it is very plausible that some of the participants had no opinion about it before the survey, hence for these participants the outcome was brought into existence and 
influenced by the context at the time the question was asked, including the specific conceptual structure of how the question was formulated. This is how context plays an essential role whenever the human mind is concerned with outcomes of cognitive tests. Like in 
physics, the probability that represents a `lack of knowledge about an actual cognitive property' is Kolmogorovian, while the probability that represents a `context-driven actualization of a cognitive property' is non-Kolmogorovian and, possibly, quantum \cite{a1986}.

The effect context plays on a conceptual entity in a typicality judgment process is equally fundamental. Consider, for example, the concept {\it Pet}, and suppose that a sample of participants are asked to rank a list of items, like {\it Dog}, {\it Cat}, {\it Squirrel}, {\it Goldfish}, {\it Parrot}, etc., as typical examples of {\it Pet}. Most of the respondents will judge items like {\it Snake} and {\it Spider} as not very typical examples of {\it Pet}, preferring to them items like {\it Dog} and {\it Cat}. But, consider now the context expressed by {\it Did you see the type of pet he has? This explains that he is a weird person}. It is very plausible that people will judge {\it Snake} and {\it Spider} as very typical examples of {\it Pet} in this context, while {\it Dog} and {\it Cat} will score a lower typicality than in the case in which {\it Pet} is considered in the absence of any context. By using a terminology that will become familiar at the end of this section, we say that, ``in a typicality measurement, the probability that the conceptual entity {\it Pet} changes its state to {\it Snake} is low in the absence of any context, while it is high in the presence of the context {\it Weird}'' \cite{ag2005a,ag2005b}. 

The impressive analogies above, between quantum and conceptual entities,  led us to look into common theoretical foundations for quantum physics and cognition, which could justify the use of a shared mathematical formalism to model microscopic and cognitive phenomena. In this respect, long-standing research exists in mathematical physics that attempts to recover the Hilbert space formalism of quantum theory from physically justified axioms, relying on well defined empirical notions, directly connected with the operations that are usually performed in a laboratory. One of the well-known approaches to the foundations of quantum physics and quantum probability is the `Geneva-Brussels approach', initiated by Jauch \cite{j1968} and Piron \cite{p1976} in Geneva, and further 
developed by the Brussels research team (see, e.g., \cite{a1999}). This research produced a formal approach, the `State Context Property' (SCoP) formalism', where any physical entity is described in terms of the basic notions of `state', `context' and `property', which arise as a consequence of concrete physical operations on macroscopic apparatuses, such as preparation and registration devices, performed in `spatiotemporal' domains, such as physical laboratories. Measurements, outcomes, state transformations 
and probabilities can then be expressed in terms of these more fundamental notions. Also, if suitable axioms are imposed on the mathematical structure underlying the SCoP formalism, the Hilbert space structure of quantum theory emerges as a unique (up to isomorphisms) mathematical representation.

When the SCoP formalism was considered, it was clear from the beginning that its validity was very general, that is, it was able to describe not only physical entities but also entities of a more abstract nature, like conceptual entities, cultural artifacts, the mind of a person, etc. \cite{a2002}. The possibility of obtaining  an `operational and realistic description' of conceptual entities, in the sense of using the SCoP formalism in cognitive domains to formalize conceptual entities in terms of states, contexts, properties, measurements and outcome probabilities, was recently re-emphasized and the essence of the approach explained in a more systematic way  
\cite{frontiersphysics}. We briefly illustrate this operational-realistic foundation of human cognition in the following.

Let us firstly consider the empirical phenomenology of cognitive psychology. Like in physics, where laboratories define precise spatiotemporal domains, we introduce `psychological laboratories' where cognitive tests, or measurements, are performed. These measurements are performed in situations that are specifically `prepared' for the tests, which includes experimental devices (if any), structured questionnaires, human participants interacting with the questionnaires in written answers, or each other, e.g., an interviewer and an interviewed. Whenever empirical data are collected from the responses of a sample of participants, a statistics of the obtained outcomes arises. These empirical facts allow identifying entities, states, contexts, measurements, outcomes, and probabilities, as follows. 

The complex of experimental procedures conceived by the experimenter, the experimental design and the cognitive effect under investigation, are typically associated with a process of preparation of the state of a conceptual entity $A$, defined by these same procedures. Hence, like in physics, the preparation procedure sets the initial state $p_A$ of the conceptual entity $A$ under study. Let us consider, for example, a questionnaire where a participant is asked to rank on a 7-point Likert scale the membership of a list of items with respect to the concept {\it Fruits}. The questionnaire defines the state $p_{Fruits}$ of the conceptual entity {\it Fruits}. It is true that cognitive situations exist where the preparation of the state of a conceptual entity is hardly controllable. Nevertheless, the state of the conceptual entity, defined by means of such a process is a `state of affairs', i.e. expresses a `reality of the conceptual entity'. In other words, once the entity is prepared in a given state, the latter is independent of any measurement, and can be confronted with the different participants in an empirical test, leading to outcome data and their statistics, exactly like in physics.

A context $e$ is an element that can provoke a change of state of the conceptual entity. For example, the concept {\it Juicy} can function as a context for the conceptual entity {\it Fruits} leading to {\it Juicy Fruits}, which can then be considered as a state of the conceptual entity {\it Fruits}. A special context is the one introduced by the measurement itself. Indeed,  when the cognitive test starts, an interaction of a cognitive nature occurs between the conceptual entity $A$ under study and a participant in the experiment, during 
which the state $p_{A}$ of the conceptual entity $A$ generally changes, being transformed into  another state $p$. Also this cognitive interaction is formalized by means of a context $e$. For example, if the participant is asked to choose the most typical example of {\it Fruits} in a list of items, containing {\it Olive}, {\it Almond}, {\it Apple}, etc., and the answer is {\it Apple}, then the initial state $p_{Fruits}$ of the conceptual entity {\it Fruits} changes to $p_{Apple}$, i.e. to  the state describing the situation ``the fruit is an apple'',  as a consequence of the contextual interaction with the participant.

The state change of a conceptual entity due to a context may be either `deterministic', hence in principle predictable under the assumption that the state before the context acts is known, or `intrinsically probabilistic', in the sense that only the probability $\mu(p,e,p_{A})$ that the  state $p_{A}$ of $A$ changes to the state $p$ is is given and is different from 1. 
In the example above on typicality judgments, the typicality of the item {\it Apple} for the concept {\it Fruits} is formalized by means of the probability $\mu(p_{Apple},e, p_{Fruits})$ of the state transition  $p_{Fruits}\to p_{Apple}$, where the context $e$ is the context of the typicality measurement.

An interesting aspect concerns the final state of a conceptual entity after a judgment or decision. As in physics, we assume the existence of a nonempty class of cognitive tests that correspond to `ideal first kind measurements', in the sense that the measurement outcome $x_k$ uniquely determines the final state $p_k$ of the conceptual entity $A$ after the measurement described by the context $e$.

The operational-realistic approach above naturally suggests the canonical quantum representation in Hilbert space, where 
a conceptual entity is associated with a Hilbert space, states are represented by Hilbert space unit vectors, measurement outcomes are represented by orthonormal (ON) basis vectors, measurement contexts are represented by spectral families of orthogonal projection operators, probabilities and state transformations are represented by the Born and L\"{u}ders rule of quantum probability, respectively.

It is worth emphasizing that in our approach states of conceptual entities describe `modes of being' of these entities, while participants act in cognitive tests as contexts for conceptual entities, changing their states. This means that states do not describe subjective beliefs of a person, or collection of persons. Subjective  beliefs are instead incorporated in the cognitive interaction between the conceptual situation and the human participants deciding on that situation. In this respect, our operational-realistic approach to human cognition departs from other quantum cognition approaches (see, e.g., \cite{bb2012,hk2013,pb2013,pnas,frontiersphysics}).

We conclude this section with an epistemological consideration. Our research investigates the validity of quantum theory as a general, unitary and coherent theory for human cognition. Our quantum theoretical models, elaborated for specific conceptual  situations and data, derive from quantum theory as a consequence of the assumptions about this general validity. As such, these models are subject to the technical and epistemological constraints of quantum theory.  In other terms, our quantum modeling rests on a `theory based approach', and should be distinguished from an `ad hoc modeling based approach', only devised to fit data. While one should be generally suspicious of models in which free parameters are added on an `ad hoc' basis to fit data, we believe that the success of our `theory based models' to reproduce different data sets constitutes in itself a convincing argument to support its advantage over traditional modeling approaches and to extend its use to more complex cognitive situations.

\section{Modeling combinations of natural concepts\label{concepts}}
We present in this section a quantum theoretical framework in complex Hilbert space to represent concepts and their combinations, showing that it explains the empirical deviations from classicality observed in conceptual categorization as genuine expressions of quantum structures \cite{a2009,abgs2013,ags2013,ag2005a,ag2005b,as2017}.

How concepts combine to form complex conceptual structures, like sentences and texts, how combinations carry new meaning, and how human reasoning processes are activated in these combinations, have always fascinated psychologists, logicians, philosophers and linguists, and have deep implications on applied disciplines, such as computer science and artificial intelligence. Many efforts were devoted to these challenges, with very few groundbreaking results.

Already at an intuitive level, defining and formalizing concepts is a complex task, which has to account for the dimension, from the `more abstract' to the `more concrete', of a concept. Saying {\it Thing} is different from expressing  the more concrete concepts of 
{\it Fruit}, {\it Apple}, or even, more specifically, {\it This Apple}, which is finally associated with a tangible object, for instance that I would be holding in this moment in my hand. Different approaches were presented to tackle the problem of defining what a concept is and, more important, to represent concepts and their combinations, like simple conjunctions and disjunctions.

The first attempt of clarifying `what a concept is' can be traced back to Aristotle and is known as the `classical', or `rule-based' view of concepts. According to the classical view, all instances of a concept share a common set of necessary and sufficient defining properties. In this view, a concept can be identified with a `set of instantiations'. However, it was already known in the 1950s that (i) in some cases, one cannot give a set of characteristics or rules defining a concept; (ii) it is often unclear whether an object is a member of a particular category; (iii) conceptual membership of an instance strongly depends on the context \cite{w1953}.

A major blow to the classical view came from Rosch's work on color \cite{r1973,r1978}. This work showed that colors do not have any particular criterial attributes or definite boundaries, and instances differ with respect to how typical they are of a concept. This led to formulation of `prototype theory', according to which concepts are organized around family resemblances, and consist of characteristic, rather than defining, features \cite{r1983}. These features are weighted in the definition of the `prototype'. Rosch showed that people rate conceptual membership as `graded', with degree of membership of an instance corresponding to conceptual distance from the prototype. Another major approach to concept theory comes from `exemplar theory', according to which a concept is represented by a set of `salient instances' of it stored in memory, rather than a set of defining or characteristic features \cite{n1988,n1992}. A third recognized approach is `theory theory', according to which concepts take the form of `mini-theories' \cite{mm1985} or schemata \cite{rn1988}, in which the causal relationships among properties are identified. More recently, other theories of concepts have also been proposed, e.g., ‘constraint theory’ \cite{ck2000}, `connectionist CONCAT model' \cite{drh2011} 
and `emergent binding model’ \cite{ts2011}.

Coming to concept representation, for years there has been an implicit assumption that classical set theoretical structures could be used to represent concepts and their combinations. Obviously, taking into account the above aspects of concepts, like `context-dependence', `vagueness' and `graded typicality', required introducing probabilistic structures and fuzzy notions, like 
the `fuzzy set representation of concepts' \cite{z1982}: for every item $X$, a concept $A$ is associated with a `membership function' $\mu(A)$ in such a way that the conjunction `$A$ and $B$' and the disjunction `$A$ or $B$' of two concepts $A$ and $B$ respectively satisfy the `minimum rule of fuzzy set conjunction' and the `maximum rule of fuzzy set disjunction', i.e.
\begin{eqnarray}
\mu(A \ {\rm and} \ B)&=&\min \Big [ \mu(A),\mu(B) \Big ] \\ \mu(A \ {\rm or} \ B)&=&\max \Big [ \mu(A),\mu(B) \Big ] 
\end{eqnarray}
This would still allow one to associate classical (fuzzy) set theoretical connectives to conceptual conjunction and disjunction. However, a whole set of experimental findings  revealed that these these fuzzy set theoretical structures are still insufficient 
to model even simple combinations of two concepts, as follows. 

(i) `Guppy effect', or `Pet-Fish problem': people judge an item like {\it Guppy} to be a very typical example of the conjunction {\it Pet-Fish}, without judging {\it Guppy} to be a typical example of either {\it Pet} or {\it Fish} \cite{os1981}.

(ii) `Over-extension and under-extension of membership weights': people judge the membership weight of the conjunction (disjunction) to be higher (lower) than the membership weight of one or both the component concepts \cite{h1988a,h1988b}.

(iii) `Borderline contradictions':  people judge a sentence like ``John is tall and John is not tall'' to be true for some borderline case “John” \cite{ap2011}.

Findings (i)--(iii) raise the so-called `combination problem', i.e. how consistently representing the combination of two (or more) concepts in terms of the representation of the component concepts.

\begin{figure} \label{hampton} 
\begin{center} 
%M \includegraphics[scale=0.15]{hampton} 
\includegraphics[scale=0.35]{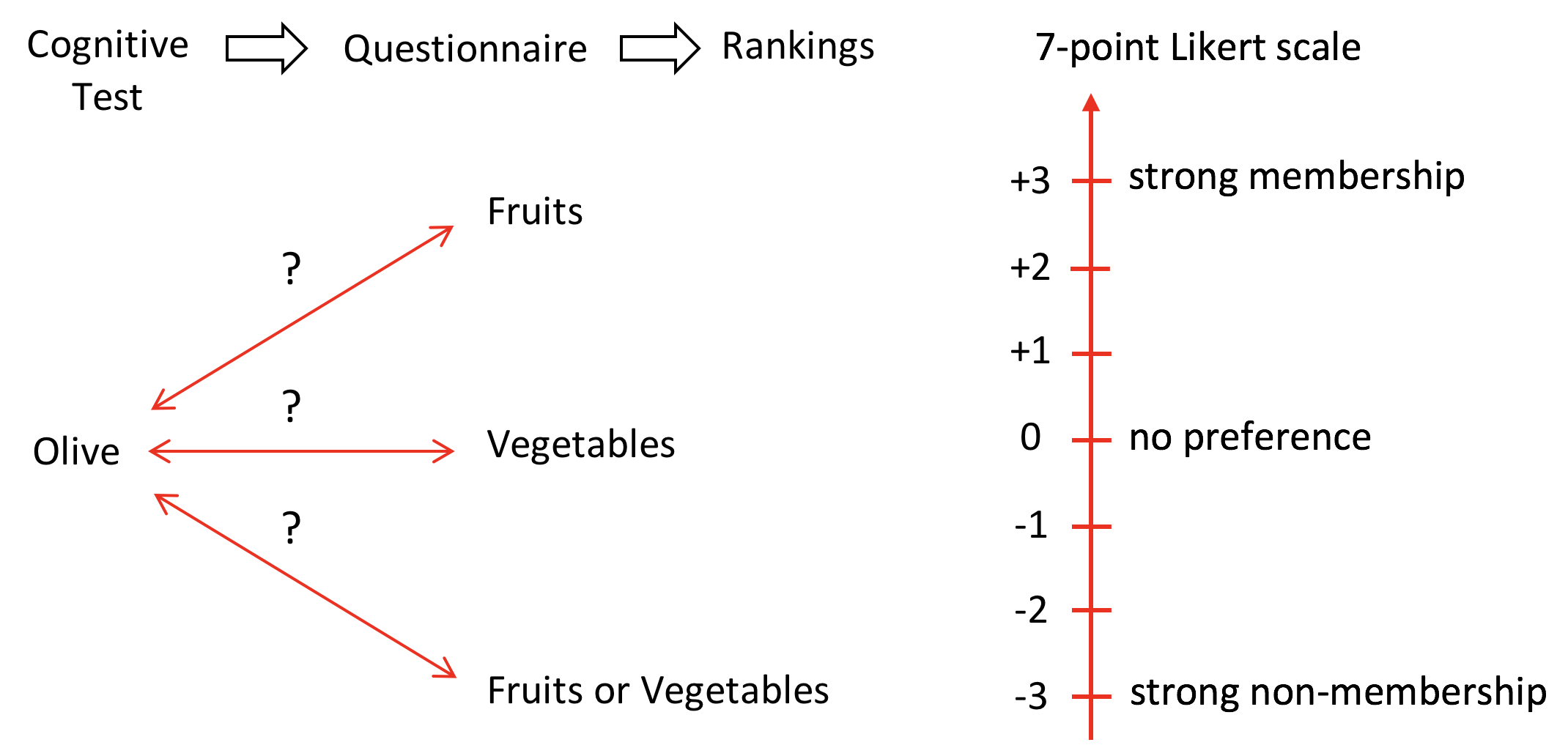} 
\end{center} {\bf Figure 1.}  {\it A schematic representation of Hampton's experiments on membership weights, where participants were asked to rank the membership of certain items, like} Olive, {\it with respect to} Fruits, Vegetables {\it and their disjunction} Fruits or Vegetables, {\it on a `7-point Likert scale'.} 
\end{figure}

Recent studies reveal that the deviation from classicality in conceptual categorization is even deeper. To this end let us analyze Hampton's experiments on membership weights. schematized in {\bf Figure 1}. Consider a pair of natural concepts, say {\it Fruits}, {\it Vegetables}, their conjunction {\it Fruits and Vegetables}, and their disjunction {\it Fruits or Vegetables}. Then, consider various items of these concepts, say {\it Apple}, {\it Tomato}, {\it Broccoli}, {\it Olive}, {\it Coconut}, {\it Mushroom}, {\it Almond}, {\it Raisin}, {\it Acorn}, etc. Next, perform a cognitive test in the form of a questionnaire submitted to a sample of participants in which they have to rank on a `7-point Likert scale' $\{+3,+2,+1,0,-1,-2,-3 \}$, membership of these items with respect to {\it Fruits}, {\it Vegetables} and their conjunction {\it Fruits and Vegetables} and 
disjunction {\it Fruits or Vegetables}. In this case, $+3$ corresponds to `strong membership', while $-3$ corresponds to `strong non-membership'. Thus, a person judging {\it Apple} as a very strong member of {\it Fruits} will write $+3$ on the questionnaire. Finally, collect the relative frequency of positive answers. In the large number limit, we call `membership weight' the obtained number.

Hampton collected membership weights of several items with respect to pairs of natural concepts and their conjunction \cite{h1988a} and/or disjunction \cite{h1988b}.
Most of Hampton's data cannot be represented by a single probability function over a $\sigma$-algebra of sets satisfying the axioms of Kolmogorov \cite{k1933}. We illustrate this important result in the following by means of general results and concrete examples.

Let us start by considering the conceptual conjunction. Let $X$ denote an item and let $\mu(A)$, $\mu(B)$ and $\mu(A \ {\rm and} \ B)$ denote the membership weights of $X$ with respect to the concepts $A$, $B$ and their conjunction `$A$ and $B$', respectively. We say that  $\mu(A)$, $\mu(B)$ and $\mu(A \ {\rm and} \ B)$ are `classical conjunction data' if a non-empty set $\Omega$, a $\sigma$-algebra $\mathscr{A}\subseteq \mathscr{P}(\Omega)$ ($\mathscr{P}(\Omega)$ denotes the power set of $\Omega$), a Kolmogorovian probability function\footnote{A normalized function $p: E \in \mathscr{A} \longrightarrow [0,1]$ is said `Kolmogorovian' if it satisfies the following axioms: (i) $p(\Omega)=1$ and (ii) $p(\cup_{i}E_i)=\sum_{i}p(E_i)$, for every sequence $\{E_i\}_i$ of pairwise disjoint events $E_i$ \cite{k1933}.} $p:E \in \mathscr{A} \longrightarrow [0,1]$, and events $E_A,E_B\in \mathscr{A}$ exist such that:
\begin{eqnarray}
\mu(A)&=&p(E_A) \\  
\mu(B)&=&p(E_B) \\ 
\mu(A \ {\rm and} \ B)&=&p(E_A \cap E_B) 
\end{eqnarray} 
One can then show that, given an item $X$, the membership weights $\mu(A)$, $\mu(B)$ and $\mu(A \ {\rm and} \ B)$ are 
`classical conjunction data' if and only if the following inequalities are simultaneously satisfied:
\begin{eqnarray}
&&\mu(A \ {\rm and} \ B)- \min \Big [ \mu(A),\mu(B) \Big ] \le 0 \label{monconj} \\
&&\mu(A)+\mu(B)- \mu(A \ {\rm and} \ B) \le 1 \label{Kolmconj}
\end{eqnarray}
Equation (\ref{monconj}) expresses monotonicity of probability with respect to the conjunction, while (\ref{Kolmconj}) is a more technical result within Kolmogorovian probability \cite{a2009}.

Most of Hampton's data on the conjunction of two concepts violate at least one of the inequalities (\ref{monconj})  and (\ref{Kolmconj}). In particular, Hampton called `overextension' a violation of (\ref{monconj}) \cite{h1988a}. Consider, for example, the item {\it Razor}. Hampton measured membership of {\it Razor} with respect to {\it Weapons}, {\it Tools} and their conjunction {\it Weapons and Tools}, finding $\mu(A)=0.63$, $\mu(B)=0.78$ and $\mu(A \ {\rm and} \ B)=0.83$, respectively. Since $0.83$ is greater than both $0.63$ and $0.78$, we get that (\ref{monconj}) is violated, and {\it Razor} is said to be `double overextended' with respect to the conjunction {\it Weapons and Tools}.

Let us now come to conceptual disjunction. Let $X$ denote an item and let $\mu(A)$, $\mu(B)$ and $\mu(A \ {\rm or} \ B)$ denote the membership weights of $X$ with respect to the concepts $A$, $B$ and their disjunction `$A$ or $B$', respectively. We say that  $\mu(A)$, $\mu(B)$ and $\mu(A \ {\rm or} \ B)$ are `classical disjunction data' if a non-empty set $\Omega$, a $\sigma$-algebra $\mathscr{A}\subseteq \mathscr{P}(\Omega)$, a Kolmogorovian probability function $p:E \in \mathscr{A} \longrightarrow [0,1]$, and events $E_A,E_B\in \mathscr{A}$ exist such that:
\begin{eqnarray}
\mu(A)&=&p(E_A) \\ 
\mu(B)&=&p(E_B) \\ 
\mu(A \ {\rm or} \ B)&=&p(E_A \cup E_B) 
\end{eqnarray}
One can then show that, given an item $X$, the membership weights $\mu(A)$, $\mu(B)$ and $\mu(A \ {\rm or} \ B)$  are `classical disjunction data' if and only if the following inequalities are simultaneously satisfied:
\begin{eqnarray}
&&\mu(A \ {\rm or} \ B)- \max \Big [ \mu(A),\mu(B) \Big ] \ge 0 \label{mondisj} \\
&&\mu(A)+\mu(B)- \mu(A \ {\rm or} \ B) \ge 0 \label{Kolmdisj}
\end{eqnarray}
Equation (\ref{mondisj}) expresses monotonicity of probability\footnote{The monotonicty law of Kolmogorovian probability is globally expressed by the inequalities $p(E_A \cap E_B)\le p(E_A),p(E_B)\le p(E_A\cup E_B)$.} with respect to the disjunction, while (\ref{Kolmdisj}) is again a more technical result within Kolmogorovian probability \cite{a2009}.

Most of Hampton's data on the disjunction of two concepts violate at least one of the inequalities 
(\ref{mondisj}) and (\ref{Kolmdisj}). In particular, Hampton called `underextension' a violation of (\ref{mondisj}) \cite{h1988b}. Consider, for example, the item {\it Ashtray}. Hampton measured membership of {\it Ashtray} with respect to {\it Home Furnishing}, {\it Furniture} and their disjunction {\it Home Furnishing or Furniture}, finding $\mu(A)=0.70$, $\mu(B)=0.30$ and $\mu(A \ {\rm or} \ B)=0.25$, respectively. Since $0.25$ is lower than both $0.70$ and $0.30$, we get that (\ref{mondisj}) is violated and {\it Ashtray} is 
said to be `double underextended' with respect to the disjunction {\it Home Furnishing or Furniture}.

The conclusion one draws from the above results is immediate. Classical (fuzzy) set theoretical logical and probabilistic structures are too limited to represent even simple combinations of two concepts.

On the other hand, a concept is not any more a set of instantiations according to the operational-realistic approach to cognition, as we have explained in Section~\ref{brussels}. A concept is rather an entity in a specific state changing under the influence of a context \cite{frontiersphysics}. We elaborated on this, developing a quantum theoretical approach to model concepts and their combinations, showing that it faithfully represents various sets of data on the Guppy effect on typicality \cite{ag2005a,ag2005b}, over- and under-extensions of membership weights \cite{a2009,as2017} and borderline contradictions \cite{s2014}. We present here the quantum theoretical model in Hilbert space for the conjunction and the disjunction of two concepts that allows  one to represent most of Hampton's data, explaining at the same time the observed deviations from classicality as consequences of genuine quantum effects, namely, `contextuality', `emergence', `interference', and `superposition'. We use the mathematical formulation and notation of quantum theory that was introduced by Paul Maurice Dirac, one of the founding fathers of quantum theory \cite{d1958}.

Let us start by the quantum model for the conjunction of two concepts. As above, let $X$ denote an item and let $\mu(A)$, $\mu(B)$ and $\mu(A \ {\rm and} \ B)$ denote the membership weights of $X$ with respect to the concepts $A$, $B$ and their conjunction `$A$ and $B$', respectively. We associate this conceptual situation with a Hilbert space ${\mathscr H}$ over the field $\mathbb{C}$ of complex numbers. Then, we represent $A$ by the unit vector $|A\rangle \in  {\mathscr H}$, i.e.  $\langle A|A\rangle=1$, and we represent $B$ by the unit vector $|B\rangle \in  {\cal H}$, i.e. $\langle B|B\rangle=1$. For the sake of simplicity, we assume that $|A\rangle $ and $ |B\rangle $ are orthogonal, i.e. the scalar product $\langle A|B\rangle =0$. Next, we represent the conjunction `$A$ and $B$' by the linear combination 
\begin{equation}
|A \ {\rm and} \ B \rangle=\frac{1}{\sqrt{2}}(|A\rangle +|B\rangle )
\label{Superpositionstate}
\end{equation}
Finally, we represent the `decision measurement' of a person judging membership of $X$ with respect to $A$, $B$ and `$A$ and $B$' by an 
orthogonal projection operator $M$ over ${\mathscr H}$, i.e. $M^{\dag}M=M$.\footnote{We remind that an orthogonal projection operator is a liner operator which satisfies hermiticity, i.e. $M^{\dag}=M$, and idempotency, i.e. $M^2=M\cdot M=M$.}  

We interpret the membership weight as a probability of membership and represent the latter by the Born rule of quantum probability \cite{d1958}. We have:
\begin{eqnarray} \mu (A)&=&\langle A|M|A\rangle \label{BornA} \\ \mu (B)&=&\langle B|M|B\rangle \label{BornB} \\  \mu(A\ \mathrm{and}\ B)&=&\langle A \ {\rm and} \ B|M| A \ {\rm and} \ B \rangle \label{Bornconj} \end{eqnarray}
By using (\ref{BornA})--(\ref{Bornconj}) and applying linearity of Hilbert space and hermiticity of $M$, i.e. $\langle B|M|A\rangle^{\ast}=\langle A|M^{\dag}|B\rangle=\langle A|M|B\rangle $ (where $\langle B|M|A\rangle^{\ast}$ is the complex conjugate of $\langle B|M|A\rangle$), we then get:
\begin{eqnarray}
\mu (A\ \mathrm{and}\ B) &=& {1\over 2}(\langle A|+\langle B|) M(|A\rangle +|B\rangle )\nonumber\\ 
&=&{\frac{1}{2}} \Big (\langle A|M|A\rangle +\langle
A|M|B\rangle +\langle B|M|A\rangle +\langle B|M|B\rangle \Big )
\nonumber \\
&=&{\frac{\mu (A)+\mu (B)}{2}}+\operatorname{Re} \, \langle A|M|B\rangle\label{intconj}
\end{eqnarray}
where $\operatorname{Re}\, \langle A|M|B\rangle$ 
is the real part of the complex number $\langle A|M|B\rangle $. We recognize in $\operatorname{Re}\, \langle A|M|B\rangle$  
the typical interference term of the quantum double-slit experiment. Its presence expresses quantum interference, as this term produces a fluctuation around the average value ${\frac{1}{2}}(\mu (A)+\mu (B))$, which is what one would classically expect in the double-slit experiment.

We now manipulate (\ref{intconj}) to get the `quantum probability formula for the conjunction', splitting it into two cases, as follows.

(I) If $\mu (A)+\mu (B)\leq 1$, we get: 
\begin{equation}
\mu (A\ \mathrm{and}\ B) ={\frac{\mu (A)+\mu (B)}{2}}+\sqrt{\mu(A)\mu(B)}\cos\theta_c
\label{probconj<1}
\end{equation}

(II) If $\mu (A)+\mu (B) > 1$, we get: 
\begin{equation}
\mu (A\ \mathrm{and}\ B) ={\frac{\mu (A)+\mu (B)}{2}}+\sqrt{1-\mu(A)}\sqrt{1-\mu(B)}\cos\theta_c
\label{probconj>1}
\end{equation}
where $\theta_c$ is the `interference angle for the conjunction'.

The quantum model above can be effectively realized in the Hilbert space ${\mathbb C}^{3}$, i.e.  the set  of all ordered triples of complex numbers.\footnote{Indeed, $|A\rangle$ and $|B\rangle$ are orthogonal vectors, and also $M|A\rangle$ and $(\mathbbmss{1}-M)|A\rangle$ and $M|B\rangle$ and $(\mathbbmss{1}-M)|B\rangle$ are, representing three data points $\mu(A)$, $\mu(B)$ and $\mu(A \ {\rm and} \ B)$ requires a Hilbert space of at least dimension 3.} Then, let $\{ (1,0,0), (0,1,0), (0,0,1)\}$ be the canonical orthonormal (ON) basis of ${\mathbb C}^{3}$. 
We again distinguish two cases, as follows.

(I) If $\mu (A)+\mu (B)\leq 1$, $\mu(A),\mu(B)\ne 0,1$, then (\ref{BornA})--(\ref{probconj<1}) are satisfied by the choice:
\begin{eqnarray}
|A\rangle &=&\Big (\sqrt{1-\mu(A)},0,\sqrt{\mu(A)} \Big ) \label{A<1}  \\
|B\rangle&=&e^{i \theta_c}\Big (-\sqrt{\frac{\mu(A)\mu(B)}{1-\mu(A)}}, -\sqrt{\frac{1-\mu(A)-\mu(B)}{1-\mu(A)}},\sqrt{\mu(B)} \Big ) \label{B<1}
\end{eqnarray}
while the orthogonal projection operator $M$ projects on the  one-dimensional 
subspace generated by $(0,0,1)$. 

(II) If 
$\mu (A)+\mu (B)> 1$, $\mu(A),\mu(B)\ne 0,1$, then (\ref{BornA})--(\ref{intconj}) and (\ref{probconj>1}) are satisfied by the choice:
\begin{eqnarray}
|A\rangle &=&\Big (\sqrt{\mu(A)},0,\sqrt{1-\mu(A)} \Big ) \label{A>1}   \\
|B\rangle&=&e^{i \theta_c}\Big (\sqrt{\frac{(1-\mu(A))(1-\mu(B))}{\mu(A)}}, \sqrt{\frac{\mu(A)+\mu(B)-1}{\mu(A)}},-\sqrt{1-\mu(B)} \Big ) \label{B>1}
\end{eqnarray}
while the orthogonal projection operator $M$ projects on the two-dimensional 
subspace generated by $(1,0,0)$ and $(0,1,0)$.\footnote{The situation in which $\mu(A)=0$ or $\mu(B)=0$ requires some further technicalities and a more complex Hilbert space structure, the `Fock space', which will be introduced later. We do not dwell on this aspect here, for the sake of brevity.}

Let us consider again the item {\it Razor} with respect to {\it Weapons}, {\it Tools} and {\it Weapons and Tools}. Since $\mu(A)+\mu(B)=0.63+0.78=1.41>1$, we use the quantum formulas in (\ref{probconj>1}), (\ref{A>1}) and (\ref{B>1}) to model the data. In particular, we get from (\ref{probconj>1}):
\begin{equation}
\theta_c=\arccos \Big ( \frac{2\mu(A \ {\rm and} \ B)-\mu(A)-\mu(B)}{\sqrt{1-\mu(A)}\sqrt{1-\mu(B)}}   \Big )=64.02^{\circ}
\label{thetac}
\end{equation}
Thus, the membership weights $\mu(A)=0.63$, $\mu(B)=0.78$ and $\mu(A \ {\rm and} \ B)=0.83$ are represented by $\theta_c=64.02^{\circ}$, $|A\rangle=(0.79,0,0.61)$ and $|B\rangle=e^{i 64.02^{\circ}}(0.36,0.81,-0.47)$ in the quantum model. In this case, double overextension of {\it Razor} can be explained as an effect of `constructive interference' ($\theta_c<90^{\circ}$) between the concepts {\it Weapons} and {\it Tools} in the superposition {\it Weapons and Tools}.

Let us now come to the quantum model for the disjunction. Again, let $X$ denote an item and let $\mu(A)$, $\mu(B)$ and $\mu(A \ {\rm or} \ B)$ denote the membership weights of $X$ with respect to the concepts $A$, $B$ and their disjunction `$A$ or $B$', respectively. We associate this conceptual situation with a complex Hilbert space ${\mathscr H}$, represent $A$ and $B$  by the orthogonal unit vectors $|A\rangle$ and $|B\rangle$, respectively, and the disjunction `$A$ or $B$' by the normalized superposition with equal weights $|A \ {\rm or} \ B \rangle=\frac{1} { \sqrt{2}}(|A\rangle +|B\rangle )$.
Finally, we represent the `decision measurement' of a person judging membership of $X$ with respect to $A$, $B$ and `$A$ or $B$' by the orthogonal projection $M$. 

By using again the Born rule of quantum probability for the membership weights $\mu(A)$ and $\mu(B)$, we find: 
\begin{equation} 
\mu(A\ \mathrm{or}\ B)=\langle A \ {\rm or} \ B|M| A \ {\rm or} \ B \rangle={\frac{\mu (A)+\mu (B)}{2}}+\operatorname{Re}\, \langle A|M|B\rangle 
\label{Borndisj} \end{equation}
As previously, (\ref{Borndisj}) mathematically takes two forms, as follows.

(I) If $\mu (A)+\mu (B)\leq 1$, we have:
\begin{equation}
\mu (A\ \mathrm{or}\ B) ={\frac{\mu (A)+\mu (B)}{2}}+\sqrt{\mu(A)\mu(B)}\cos\theta_d
\label{probdisj<1}
\end{equation}

(II) If $\mu (A)+\mu (B) > 1$, we have:
\begin{equation}
\mu (A\ \mathrm{or}\ B) ={\frac{\mu (A)+\mu (B)}{2}}+\sqrt{1-\mu(A)}\sqrt{1-\mu(B)}\cos\theta_d
\label{probdisj>1}
\end{equation}
where $\theta_d$ denotes the `interference angle for the disjunction'. Also here, the quantum model can be realized in the Hilbert space ${\mathbb C}^{3}$, and formulas analogous to (\ref{A<1})--(\ref{B>1}) hold with $\theta_d$ in place of $\theta_c$ \cite{a2009}.

Let us consider again the item {\it Ashtray} with respect to {\it Home Furnishing}, {\it Furniture} and {\it Home Furnishing or Furniture}. Since $\mu(A)+\mu(B)=0.70+0.30=1$, we use the quantum formulas (\ref{probdisj<1}), (\ref{A<1}) and (\ref{B<1}), with $\theta_d$ in place of $\theta_c$, to model the data. In particular, we get from (\ref{probdisj<1}):
\begin{equation}
\theta_d=\arccos \Big ( \frac{2\mu(A \ {\rm or} \ B)-\mu(A)-\mu(B)}{\sqrt{\mu(A)\mu(B)}}   \Big )=123.06^{\circ}
\end{equation}
Thus, the membership weights $\mu(A)=0.70$, $\mu(B)=0.30$ and $\mu(A \ {\rm or} \ B)=0.25$ are represented by $\theta_d=123.06^{\circ}$, $|A\rangle=(0.55,0,0.84)$ and $|B\rangle=e^{i 123.06^{\circ}}(-0.84,0,-0.55)$ in the quantum model. Hence, double underextension of {\it Ashtray} can be explained as an effect of `destructive interference' ($90^{\circ}<\theta_d<180^{\circ}$) between the concepts {\it Home Furnishing} and {\it Furniture} in the superposition {\it Home Furnishing or Furniture}.

The quantum probability formulas (\ref{intconj}) and (\ref{Borndisj}) enable representation of most
of Hampton's data on conjunctions/disjunctions of 2 concepts. We report some relevant cases in {\bf Table 1}. 
\begin{table}
\begin{center}
\begin{tabular}{|cccccc|}
\hline\hline 
\multicolumn{6}{|c|}{\rule{0pt}{3ex} Experiment by Hampton on conceptual conjunction \cite{h1988a}} \\
\hline\hline
	\rule{0pt}{3ex} $\mu(A)$	&	$\mu(B)$	&	$\mu(A \ {\rm and} \ B)$ & $\theta_{c}$	 & $|A\rangle$ & $e^{-i \theta_{c}}|B\rangle$	\\
\hline
	\rule{0pt}{3ex} $0.725$	&	$0.825$	&	$0.825$	&	$76.83^{\circ}$	&	$(0.85,	0,	0.52)$	&	$(0.26,	0.87,	-0.42)$	\\
\multicolumn{6}{|c|}{$X$={\it Desk Lamp}\quad $A=${\it Furniture} \quad $B=${\it Household Appliances}} \\
\hline
	\rule{0pt}{3ex} $0.87$	&	$0.81$	&	$0.9$	&	$67.56^{\circ}$	&	$(0.93,	0,	0.36)$	&	$(0.17,	0.88,	-0.44)$	\\
\multicolumn{6}{|c|}{$X$={\it Mint}\quad $A=${\it Food} \quad $B=${\it Plant}} \\
\hline
	\rule{0pt}{3ex} $0.5$	&	$0.9$	&	$0.95$	&	$77.02^{\circ}$	&	$(0.88,	0,	0.48)$	&	$(0.21,	0.89, -0.39)$	\\
\multicolumn{6}{|c|}{$X$={\it Tree House}\quad $A=${\it Building} \quad $B=${\it Dwelling}} \\
\hline\hline
\multicolumn{6}{|c|}{\rule{0pt}{3ex} Experiment by Hampton on conceptual disjunction \cite{h1988b}} \\
\hline\hline
		\rule{0pt}{3ex} $\mu(A)$	&	$\mu(B)$	&	$\mu(A \ {\rm or} \ B)$ & $\theta_{d}$	 & $|A\rangle$ & $e^{-i \theta_{d}}|B\rangle$	\\
\hline
	\rule{0pt}{3ex} $0.5$	&	$0.7$	&	$0.4$	&	$121.09^{\circ}$	&	$(0.71,	0,	0.71)$	&	$(0.55,	0.63,	-0.55)$	\\
\multicolumn{6}{|c|}{$X$={\it Rat}\quad $A=${\it Pets} \quad $B=${\it Farmyard Animals}} \\
\hline
	\rule{0pt}{3ex} $0.7$	&	$0.7$	&	$1$	&	$121.90^{\circ}$	&	$(0.73,	0,	0.68)$	&	$(0.61,	0.45, -0.66)$	\\
\multicolumn{6}{|c|}{$X$={\it Tomato}\quad  $A=${\it Fruits} \quad $B=${\it Vegetables}} \\
\hline
	\rule{0pt}{3ex} $0.4$	&	$0.7$	&	$0.95$	&	$53.13^{\circ}$	&	$(0.71,	0,	0.71)$	&	$(0.71,	0,	-0.71)$	\\
\multicolumn{6}{|c|}{$X$={\it Cake Tin}\quad $A=${\it Household Appliances} \quad $B=${\it Kitchen Utensils}} \\
\hline
\end{tabular}
\end{center}
{\bf Table 1.} {\it Quantum modeling of some of Hampton's data on the conjunction and the disjunction of two concepts.} 
\end{table}
It is worth observing, at this stage, that the quantum theoretical model in the Hilbert space $\mathscr{H}$ is not able to fully represent conjunctions and disjunctions of two concepts in all their generality. Indeed, a more general structure is needed in this case, because of reasons that will become 
clear in Section \ref{explanation}. This more general algebraic structure is a `two-sector Fock space' $\mathscr{F}$, i.e. $\mathscr{F}=\mathscr{H} \oplus (\mathscr{H} \otimes \mathscr{H})$ ($\oplus$ denotes the direct sum of linear vector spaces), where the individual Hilbert space $\mathscr H$ is called the 
`sector 1 of $\mathscr{F}$', while the tensor product Hilbert space  $\mathscr{H} \otimes \mathscr{H}$ is 
called the 
`sector 2 of $\mathscr{F}$' \cite{a2009,as2017,s2014,s2015}.

The Fock space representation explains conceptual phenomena that are much deeper than the ones observed above. To understand what we mean by this, let us study a relevant example, as follows.

Hampton measured membership weights of the item {\it Olive} with respect to the concepts {\it Fruits}, {\it Vegetables} and their disjunction {\it Fruits or Vegetables}, finding 0.5, 0.1 and 0.8, respectively \cite{h1988b}. In this case, (\ref{Kolmdisj}) is violated, hence these data are non-classical. Now, consider again {\it Olive} and its membership weights with respect to the concepts {\it Fruits}, {\it Vegetables} and their conjunction {\it Fruits and Vegetables}. We performed the test, finding 0.56, 0.63 and 0.65, respectively \cite{s2015}. In this case, (\ref{monconj}) is violated, hence these data are again non-classical. However, we notice that in both the disjunction and the conjunction, {\it Olive} is double overextended, with similar weights for the combined concept. This clearly indicates that situations exist where people do not really take into account whether the connective `or' or the connective `and' is considered, but they actually estimate whether the given item is a member of the new emergent concept, obtained by combining two concepts, but in a way that is independent of the fact that the combination is realized through a disjunction or a conjunction. 
An intuitive and general notion of `conceptual emergence' thus arises, which will be precisely defined in Section \ref{explanation}. 

The quantum theoretical approach in Fock space enables discovery of new non-classical effects. In this regards, we recently generalized Hampton's experiments, extending them to conjunctions and negations of two natural concepts. We tested in an experiment \cite{s2015} membership weights of items with respect to:

(i) natural concepts $A$ and $B$;

(ii) conceptual negations `not $A$', `not $B$';

(iii) conjunctions `$A$ and $B$', `$A$ and not $B$', `not $A$ and $B$', and `not $A$ and not $B$'.

As expected, we identified systematic (double) overextension in all the conjunctions above, as well as deviations from classicality in conceptual negations. But, we also found a new non-classical effect \cite{asv2015,PhilTransA2015}. Indeed we identified unexpected, systematic and significant deviations from the marginal law. Consider, for example, the quantity
\begin{equation}
\mu(A \ {\rm and } \ B)+\mu(A \ {\rm and } \ {\rm not} \ B)-\mu(A) \label{marginal}
\end{equation}
In a Kolmogorovian probability framework, (\ref{marginal}) should be identically zero, for every item and pair of concepts. Since we know that membership weight data violate Kolmogorovian modeling, one would expect that the left side of (\ref{marginal}) is different from 0 and generally depends on the item and the pair of concepts that are tested. We instead found that the following equality is 
approximately satisfied
\begin{equation}
\mu(A \ {\rm and } \ B)+\mu(A \ {\rm and } \ {\rm not} \ B)-\mu(A) \approx \frac{1}{2} \label{qmarginal}
\end{equation}
In addition, the numerical value on the right side of (\ref{qmarginal}) does not depend on either the item or the pair of concepts 
that are tested. This non-classical pattern is predicted by our quantum theoretical model in Fock space for the conjunction and negation of two natural concepts. More precisely, it can be shown that the left hand side of (\ref{qmarginal}) is given by interference terms oscillating around the value 0.5, which will generally compensate each other. Thus, observing that emergent reasoning (represented by sector 1 of Fock space) typically prevails over logical reasoning (represented by sector 2 of Fock space), as confirmed by the experimental data, it follows that the approximate equality (\ref{qmarginal}) will hold in general  \cite{asv2015,PhilTransA2015}. 
This means that our quantum model is able to capture a robust non-classical effect which goes much deeper than just over- and under-extension, but is part of an intrinsic mechanism of concept formation \cite{as2017,PhilTransA2015,IQSA2}.

Hence, the quantum theoretical approach for conceptual entities and their combinations produces highly predictive models, which have allowed us to identify further genuine quantum effects in the combination of concepts. For example, we 
identified an experimental violation of classical Bell's inequalities for the concept combination {\it The Animal Acts}, and elaborated a quantum representation for it, thus proving that the violation can be explained in terms of `quantum entanglement' \cite{as2011,as2014}. We also detected `quantum indistinguishability of Bose-Einstein type' in specific combinations of identical concepts, e.g., {\it Eleven Animals} \cite{IQSA1}. We will not go into detail here about the Fock space modeling of conceptual combinations, nor analyse how entanglement and indistinguishability were identified. We instead limit ourselves to the `one sector' Hilbert space model, and this mainly for three reasons. Firstly, because of its mathematical simplicity and tractability. Secondly, because Hilbert space representations already enable to grasp essential aspects of quantum structures in conceptual combinations (see, for example, 
%M {\bf Figure 3}, 
{\bf Figure 2}, 
where we illustrate an impressive effect of quantum interference we identified in the disjunction of {\it Fruits and Vegetables} 
%M 
\cite{Aerts2009a}.). 
Thirdly, because non-classical phenomena identified in other areas of cognitive science can be equally 
realized in a 
Hilbert space $\mathbb{C}^3$, as we will see in Sections \ref{linda} and \ref{twogamble}.
\begin{figure} \label{disjunction} 
\begin{center} 
\includegraphics[scale=0.5]{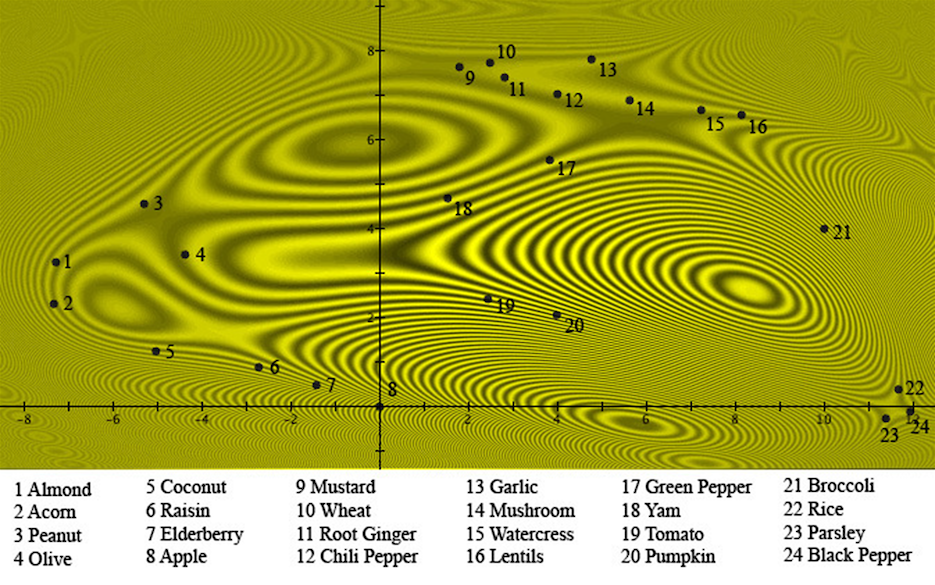} 
\end{center} 
%M {\bf Figure 3.} 
{\bf Figure 2.} 
{\it Quantum interference effects in the disjunction} Fruits or Vegetables {\it of the natural concepts} Fruits {\it and} Vegetables. {\it The typical interference fringes of the double-slit experiment are recognizable. 
%M 
For more details about how this figure was obtained, see for instance \cite{Aerts2009a}.} 
\end{figure}

\section{Application to conjunctive fallacies\label{linda}}
Non-classical effects have been identified in different domains of cognitive science, as mentioned in Section~\ref{intro}. 
In this section, we apply the quantum conceptual approach presented in Section \ref{concepts} to model conjunctive and disjunctive fallacies, which are examples of probability judgment errors.

In 1983, Kahneman and Tversky identified an important deviation of classicality in decision science, which is known as the `conjunction fallacy' \cite{tk1983,bpft2011,bb2012}. More precisely, Kahneman and Tversky performed a cognitive test on a sample of participants, presenting them a questionnaire containing a story about a liberal woman named `Linda', as follows.

``Linda is 31 years old, single, outspoken and very bright. She majored in philosophy. As a student, she was deeply concerned with issues of discrimination and social justice, and also participated in anti-nuclear demonstrations.''

Participants were then interrogated about the likelihood of the following events:

(i) Linda is a bank teller;

(ii) Linda is a bank teller and is active in the feminist movement.

Kahneman and Tversky found that 85\% of the respondents judged option (ii) to be more likely than option (i). Obviously, 
the test violates monotonicity of classical probability, and such violation exactly expresses the conjunction fallacy \cite{tk1983}.

Several empirical studies have confirmed the conjunction fallacy, and various competing proposals have been formulated at a theoretical level (for a literarary review, see, for example \cite{mo2009}). As mentioned in Section~\ref{intro}, an 
influential proposal of solution came from Kahneman and Tversky themselves within their theory of `individual heuristics and biases' \cite{tk1974}. They formulated the `judgment heuristics hypothesis', according to which people prefer the judgment heuristics of `representativeness' to standard logical reasoning, and option (ii) is more representative than option (i) of Linda's activities. An alternative proposal is the `misunderstanding hypothesis': people misunderstand some terms of the problem, like `and' and `probability'. More recently, some authors have put forward that the conjunction fallacy is a consequence of `question order effects', that is, people first judge the most likely event (i) and then the event (ii), where (i) and (ii) are incompatible in the standard quantum sense \cite{bpft2011,bb2012}.

Morier and Borgida performed a more complete test of the fallacy in 1984 \cite{mb1984}. They asked a sample of participants to rank the likelihood of the following events:

(i) Linda is a feminist;

(ii) Linda is a bank teller;

(iii) Linda is a feminist and a bank teller;

(iv) Linda is a feminist or a bank teller.

Let us denote by $\mu(A)$, $\mu(B)$, $\mu(A \ {\rm and} \ B)$ and $\mu(A \ {\rm or} \ B)$ the mean probability that Linda is judged as a feminist, bank teller, feminist and bank teller, and feminist or bank teller, respectively. Morier and Borgida found the following mean 
probability judgments'
\begin{eqnarray} \mu(A)=0.83&>&0.60=\mu(A \ {\rm or} \ B) \label{disjunctionfallacy} \\ \mu(A \ {\rm and} \ B)=0.36&>&0.26=\mu(B) \label{conjunctionfallacy} \end{eqnarray}
Equation (\ref{conjunctionfallacy}) expresses a conjunction fallacy, because the probability of the conjunction is judged as higher than the probability of one of the two events. Equation (\ref{disjunctionfallacy}) expresses instead  a `disjunction fallacy', because the probability of the disjunction is judged as lower than the probability of one of the two events \cite{mb1984}.

By referring to the terminology introduced in Section~\ref{concepts}, we recognize strong similarities between conjunction fallacy and conceptual overextension on the one side, and between disjunction fallacy and conceptual underextension on the other side. And, indeed, the quantum theoretical model of conceptual conjunction and disjunction can be successfully applied to also model conjunctive and disjunctive fallacies, respectively.

Let us start by the conjunction fallacy and let us refer to the quantum representation of the conjunction of two concepts in Section~\ref{concepts}. We denote the concept {\it Feminist} by $A$ and the concept {\it Bank Teller} by $B$. The conjunction `$A$ and $B$' then corresponds to the conceptual conjunction {\it Feminist and Bank Teller}. As above, we denote by $\mu(A)$, $\mu(B)$ and $\mu(A \ {\rm and} \ B)$ the mean probability that the item {\it Linda} is judged to be a {\it Feminist}, a {\it Bank Teller} and a {\it Feminist and Bank Teller}, respectively. These probabilities can be interpreted as membership weights of the item {\it Linda} with respect to the concepts {\it Feminist}, {\it Bank Teller} and {\it Feminist and Bank Teller}, respectively, in accordance with the analysis in Section~\ref{concepts}.

We associate the overall conceptual situation above with a Hilbert space ${\mathscr H}$ over complex numbers. Then, as in Section~\ref{concepts}, we represent $A$ and $B$ by the unit vectors $|A\rangle $ and $|B\rangle $, respectively, of $\mathscr{H}$. We again assume that $|A\rangle $ and $ |B\rangle $ are orthogonal, and represent the conjunction `$A$ and $B$' by the normalized superposition state vector $\frac{1}{\sqrt{2}}(|A\rangle+|B\rangle)$, as in(\ref{Superpositionstate}). Finally, the decision measurement of a person judging whether Linda is `a feminist', `a bank teller', and `a feminist and a bank teller' is represented by the orthogonal projection operator $M$ over ${\mathscr H}$.

By using the Born rule of quantum probability, we  can now write (\ref{BornA}), (\ref{BornB}), (\ref{Bornconj}), where the latter,  
%M becomes: 
as we observed already, can be written in the form:
\begin{equation} 
\label{probconjfall} 
\mu(A\ \mathrm{and}\ B)=\left\{ 
\begin{array}{lcc} {\frac{\mu (A)+\mu (B)}{2}}+\sqrt{\mu(A)\mu(B)}
%M \cos\theta_d 
\cos\theta_c 
& {\rm if} & \mu(A)+\mu(B)\le 1 \\ {\frac{\mu (A)+\mu (B)}{2}}+\sqrt{1-\mu(A)}\sqrt{1-\mu(B)}
%M \cos\theta_d 
\cos\theta_c 
& {\rm if} & \mu(A)+\mu(B)> 1 \\ 
\end{array}  \right. 
\end{equation}
This completes the construction of a quantum theoretic model for the conjunction fallacy. One easily shows that the quantum model can be realized in the Hilbert space $\mathbb{C}^3$, and (\ref{A<1}) and (\ref{B<1}) hold in case (I), while (\ref{A>1}) and (\ref{B>1}) hold in case (II), Section \ref{concepts}.

Let us now show that the quantum model faithfully represents conjunctive data in \cite{mb1984}. In this case, the mean judgment probabilities are $\mu(A) = 0.83$, $\mu(B) = 0.26$ and $\mu(A\ {\rm and}\ B) = 0.36$. Since $\mu (A)+\mu (B)=1.09>1$, we get from 
(\ref{A>1}), (\ref{B>1}) and (\ref{probconjfall}):
\begin{equation}
\theta_c=\arccos \Big ( \frac{2\mu(A \ 
%M {\rm or} 
 {\rm and} 
\ B)-\mu(A)-\mu(B)}{\sqrt{1-\mu(A)}\sqrt{1-\mu(B)}}   \Big )=121.44^{\circ}
\end{equation}
\begin{eqnarray} |A\rangle&=&(0.91,0,0.41) \\ |B\rangle&=&e^{i 121.44^{\circ}}(0.39,0.33,-0.86) \end{eqnarray}
in the canonical basis of $\mathbb{C}^{3}$. The explanation we give to the modeling above is that, when the item {\it Linda} is considered, together with her story,  the concepts {\it Feminist} and {\it Bank Teller} `destructively interfere' in the conjunction {\it Feminist and Bank Teller}, meant as a newly emerging conceptual entity.

We can also provide a quantum representation for the disjunction fallacy in \cite{mb1984}. We now let the superposition vector $\frac{1}{\sqrt{2}}(|A\rangle+|B\rangle)$  represent the conceptual disjunction {\it Feminist or Bank Teller}. Then, 
(\ref{A<1}) (or (\ref{A>1})), (\ref{B<1}) (or (\ref{B>1})) and (\ref{probconjfall}) 
hold with $\theta_d$ in place of $\theta_c$. In \cite{mb1984}, the mean judgment probabilities are $\mu(A) = 0.83$, $\mu(B) = 0.26$ and $\mu(A\ {\rm or}\ B) = 0.60$. Since $\mu (A)+\mu (B)=1.09>1$, we get:
\begin{equation}
\theta_d=\arccos \Big ( \frac{2\mu(A \ {\rm or} \ B)-\mu(A)-\mu(B)}{\sqrt{1-\mu(A)}\sqrt{1-\mu(B)}}   \Big )=81.08^{\circ}
\end{equation}
\begin{eqnarray} |A\rangle&=&(0.91,0,0.41) \\ |B\rangle&=&e^{i 81.08^{\circ}}(0.39,0.33.-0.86) \end{eqnarray}
in the canonical basis of $\mathbb{C}^{3}$. As we can see, the quantum model of conceptual disjunction faithfully represents data on the disjunction fallacy. The explanation we give to the modeling above is that, when the item {\it Linda} is considered, together with her story,  the concepts {\it Feminist} and {\it Bank Teller} `constructively interfere' in the disjunction {\it Feminist or Bank Teller}, meant as a newly emerging conceptual entity.

Several tests on Linda-like stories have been performed in the last thirty years, finding systematic deviations from classicality. We report in {\bf Table 2} some of these data, together with the corresponding values from the quantum model. For each experimental test, we take the average of the judgment probabilities across the various Linda-like stories.
\begin{table}
\begin{center}
\begin{tabular}{|ccccccc|}
\hline\hline
\multicolumn{7}{|c|}{\rule{0pt}{3ex} Tests on the conjunction fallacy} \\
\hline\hline
\rule{0pt}{3ex} {\it Experiment}	&	$\mu(A)$	&	$\mu(B)$	&	$\mu(A \ {\rm and} \ B)$ & $\theta_{c}$	 & $|A\rangle$ & $e^{-i \theta_{c}}|B\rangle$	\\
\hline
\rule{0pt}{3ex}
{T \& K (1983)}	&	$0.61$	&	$0.38$	&	$0.51$	&	$88.21^{\circ}$	&	$(0.62,	0,	0.78)$	&	$(-0.77,	-0.16,	0.62)$	\\
\hline
\rule{0pt}{3ex}
{F \& P (1996)}	&	$0.61$	&	$0.37$	&	$0.39$	&	$102.15^{\circ}$	&	$(0.62,	0,	0.78)$	&	$(-0.76,	-0.23,	0.61)$	\\
\hline
\rule{0pt}{3ex}
{Lu (2015)}	&	$0.59$	&	$0.32$	&	$0.37$	&	$101.28^{\circ}$	&	$(0.64,	0,	0.77)$	&	$(-0.68,	-0.47, 0.57)$	\\
\hline\hline
\multicolumn{7}{|c|}{\rule{0pt}{3ex} Tests on the disjunction fallacy} \\
\hline\hline
\rule{0pt}{3ex} {\it Experiment}	&	$\mu(A)$	&	$\mu(B)$	&	$\mu(A \ {\rm or} \ B)$ & $\theta_{d}$	 & $|A\rangle$ & $e^{-i \theta_{d}}|B\rangle$	\\
\hline
\rule{0pt}{3ex}
{Fisk (2002)}	&	$0.63$	&	$0.33$	&	$0.54$	&	$82.44^{\circ}$	&	$(0.61,	0,	0.79)$	&	$(-0.75,	-0.33,	0.57)$	\\
\hline
\rule{0pt}{3ex}
{Lu (2015)}	&	$0.57$	&	$0.26$	&	$0.54$	&	$71.05^{\circ}$	&	$(0.66,	0,	0.75)$	&	$(-0.59,	-0.63,	0.51)$	\\
\hline
\end{tabular}
\end{center}
{\bf Table 2.} {\it Quantum modeling of mean probability data averaged across different Linda-like tests in Tversky and Kahneman (T \& K) \cite{tk1983}, Fisk and Pidgeon (F \& P) \cite{fp1996} and Lu  \cite{lu2015} for the conjunction fallacy, and Fisk \cite{f2002} and Lu \cite{lu2015} for the disjunction fallacy.} 
\end{table}
To conclude the section, conjunction and disjunction fallacies can be respectively interpreted as the decision theory counterparts of overextension and underextension of concept theory. The quantum effects of contextuality, interference and superposition can again naturally account for the observed deviations from classicality in these human reasoning fallacies. In analogy with the case of double overextension in conceptual conjunctions, the quantum model also predicts the presence of `double conjunction fallacies', though it predicts that double conjunction fallacies will be less likely than single conjunction fallacies. Similarly, the quantum model predicts the presence of `double disjunction fallacies', in analogy with the case of double underextension in conceptual disjunctions.

\section{Application to disjunctive effects\label{twogamble}}
We have seen in Section \ref{linda} that violations of classical logical reasoning occur in concrete human decisions in the form of probability judgement errors. In this section, we want to show 
that pitfalls of classical decision theory are also observed in the form of decision-making errors.

As mentioned in Section \ref{intro}, EUT is the predominant model of decision-making under uncertainty: it prescribes that in the presence of uncertainty, decision makers should choose in such a way to maximize their utility, or degree of satisfaction \cite{s1954}. However, the `Ellsberg' \cite{e1961} and `Machina paradoxes' \cite{m2009} reveal that cognitive factors play a fundamental role in concrete decision-making processes which are not accounted for by EUT. One of the building axioms of EUT, Savage's `sure thing principle', is specifically violated in the Ellsberg and Machina paradoxes. We extensively studied in \cite{ahs2017,as2016,agms2018} how the quantum theoretical approach solves these paradoxical situations. For sake of brevity, we do not deal with them here, but we instead discuss the violation of the sure thing principle that is observed in other decision-making experiments.

Savage stated the sure thing principle being inspired by the following story \cite{s1954}.

``A businessman contemplates buying a certain piece of property. He considers the outcome of the next presidential election relevant. So, to clarify the matter to himself, he asks whether he would buy if he knew that the Democratic candidate were going to win, and decides that he would. Similarly, he considers whether he would buy if he knew that the Republican candidate were going to win, and again finds that he would. Seeing that he would buy in either event, he decides that he should buy, even though he does not know which event obtains, or will obtain, as we would ordinarily say.'' 

Tversky and Shafir tested the sure thing principle in an experiment where they presented a group of students with a `two-stage gamble', that is, a gamble which can be played two sequential times \cite{bb2012,ts1992}. At each stage the decision consisted in whether or not playing a gamble that has an equal chance of winning, say $\$200$, or losing, say $\$100$. The key result is based on the decision for the second bet, after finishing the first bet. The experiment included three situations: 

(i) the students were informed that they had already won the first gamble; 

(ii) the students were informed that they had lost the first gamble; 

(iii) the students did not know the outcome of the first gamble. 

Tversky and Shafir found that 69\%, that is, the majority of the students who knew they had won the first gamble, decided to play again; 
59\%, again the majority of the students who knew they had lost the first gamble, decided to play again; but only 36\% of the students who did not know whether they had won or lost chose to play again (equivalently, 64\%, that is, the majority, decided not to play again) \cite{ts1992}. 

The two-stage gamble experiment violates Savage's sure thing principle. Indeed, students generally decide to play again if they know they won, and they also decide to play again if they know they lost, but they generally decide not to play again when they do not know whether they won or lost.

More generally, the experiment performed by Tversky and Shafir violates the total law of Kolmogorovian probability.\footnote{In Kolmogorovian probability, one proves the law of total probability, namely, $p(E_A)=p(E_B)p(E_A|E_B)+p(E'_{B})p(E_A|E'_B)$, where $E'_B=\Omega\setminus E_B$ denotes the `complement event' with respect to $E_B$.} Indeed, if we denote by $\mu(P)$ the total probability that a student decides to play again without knowing whether s/he 
has won or lost the first gamble, by $\mu(W)$ and $\mu(L)$ the probability that the student wins and loses the first gamble, respectively, by $\mu(P|W)$ the conditional probability that the student decides to play again knowing s/he 
has won, and  by $\mu(P|L)$ the conditional probability that the student decides to play again knowing s/he 
has lost, then it is not possible to find any value of $\mu(W)$ and $\mu(L)=1-\mu(W)$ such that $\mu(P|W)=0.69$ and $\mu(P|L)=0.59$, $p(P)=0.36$ and the law of total probability
\begin{equation}
\mu(P)=\mu(W)\mu(P|W)+\mu(L)\mu(P|L)
\end{equation}
is satisfied. This violation of the laws of Kolmogorovian probability is called the `disjunction effect'. An equivalent formulation of the disjunction effect is known as the `Hawaii problem', and it is again due to Tversky and Shafir \cite{ts1992}, while a disjunction effect also occurs in the `prisoner's dilemma' \cite{bb2012}. A seemingly plausible explanation is that the origin of the violation of the sure thing principle in the disjunction effect is `uncertainty aversion', that is, people prefer certain over uncertain events.

Let us now work out a quantum theoretical model for the disjunction effect. Like in the disjunction fallacy, we need to suitably apply the quantum modelof Section \ref{concepts} about the disjunction of two concepts. To this end, let us denote by $A$ the conceptual situation of `having won the first gamble', by $B$ the conceptual situation of `having lost the first gamble', and by `$A$ or $B$' the conceptual situation of `having won or lost the first gamble', and we denote by $\mu(A)$, $\mu(B)$ and $\mu(A \ {\rm or} \ B)$, respectively, the corresponding probabilities. The participant (student) has to make a decision whether to play again -- positive outcome, or not to play again -- negative outcome.

We associate the overall conceptual situation above with a Hilbert space ${\mathscr H}$ over complex numbers. Then, as in Sections \ref{concepts} and \ref{linda}, we represent $A$ by the unit vector $|A\rangle $ and $B$ by the unit vector $|B\rangle $ of ${\mathscr H}$. We again assume that $|A\rangle $ and $ |B\rangle $ are orthogonal and represent the disjunction `$A$ or $B$' by means of 
their normalized superposition vector $|A \ {\rm or} \ B\rangle=\frac{1}{\sqrt{2}}(|A\rangle +|B\rangle )$. 
The decision measurement has two outcomes, say `yes' and `no', hence it is represented by the orthogonal projection operator $M$ over ${\mathscr H}$. In the two-stage gamble experiment by Tversky and Shafir, we have that the probability of the outcome `yes', that is, `play again', in the `win' situation, that is, state vector $|A\rangle$, is 0.69, hence we write $\mu(A) = 0.69$. The probability of the outcome `yes', that is, `play again', in the `lost' situation, that is, the state vector $|B\rangle$, is 0.59, hence we write $\mu(B) = 0.59$. The probability of the outcome `yes', that is, `play again', in the `win or lost' situation, that is, the superposition 
state vector $|A \ {\rm or} \ B\rangle=\frac{1}{ \sqrt{2}}(|A\rangle +|B\rangle )$, is 0.36, hence we write $\mu(A\ {\rm or}\ B) = 0.36$. 

In accordance with the Born rule of quantum probability, 
%M we thus have: 
formulae (\ref{Borndisj})-(\ref{probdisj>1}) apply again, so we can write:
\begin{eqnarray} 
\mu (A)&=&\langle A|M|A\rangle \\ \mu (B)&=&\langle B|M|B\rangle \\  \mu(A\ \mathrm{or}\ B)&=&\langle A \ {\rm or} \ B|M|A \ {\rm or} \ B\rangle={\frac{\mu (A)+\mu (B)}{2}}+\operatorname{Re} \, \langle A|M|B\rangle \nonumber \\ &=&\left\{ \begin{array}{lcc} {\frac{\mu (A)+\mu (B)}{2}}+\sqrt{\mu(A)\mu(B)}\cos\theta_d & {\rm if} & \mu(A)+\mu(B)\le 1 \\ {\frac{\mu (A)+\mu (B)}{2}}+\sqrt{1-\mu(A)}\sqrt{1-\mu(B)}\cos\theta_d & {\rm if} & \mu(A)+\mu(B)> 1 \\ \end{array}  \right. \label{probdisjeffect} 
\end{eqnarray}
Equation (\ref{probdisjeffect}) includes both cases (I) and (II) in Section \ref{concepts} for the disjunction of two concepts. In particular, since $\mu(A)+\mu(B)=0.69+0.59=1.28>1$ in the Tversky-Shafir's experiment we are considering \cite{ts1992}, coming to the $\mathbb{C}^{3}$ realization, we get: 
Since in the Tversky-Shafir's experiment we are considering \cite{ts1992}, we have $\mu(A)+\mu(B)=0.69+0.59=1.28>1$, it is (\ref{probdisj>1}) that we have to use, so coming to the $\mathbb{C}^{3}$ realization, we get: 
\begin{equation}
\theta_d=\arccos \Big ( \frac{2\mu(A \ {\rm or} \ B)-\mu(A)-\mu(B)}{\sqrt{1-\mu(A)}\sqrt{1-\mu(B)}}   \Big )=141.76^{\circ}
\end{equation}
with the coordinates of two vectors $|A\rangle$ and  $|B\rangle$, in the canonical basis of $\mathbb{C}^{3}$, being given by: 
\begin{eqnarray}
|A\rangle &=&\Big (\sqrt{\mu(A)},0,\sqrt{1-\mu(A)} \Big )=(0.83,0,0.56)   \\
|B\rangle&=&e^{i \theta_d}\Big (\sqrt{\frac{(1-\mu(A))(1-\mu(B))}{\mu(A)}}, \sqrt{\frac{\mu(A)+\mu(B)-1}{\mu(A)}},-\sqrt{1-\mu(B)} \Big ) \nonumber 
\\&=&e^{i 141.76^{\circ}}(0.43,0.64,-0.64)
\end{eqnarray}
and the orthogonal projection operator $M$ projecting on the subspace generated by $(1,0,0)$ and $(0,1,0)$.

It is worth reminding that, also in this disjunction effect situation, we have applied two key quantum features, namely, superposition, in the linear combination ${\frac{1}{\sqrt{2}}}(|A\rangle +|B\rangle )$ to represent `$A$ or $B$', and interference, as the effect appearing in (\ref{probdisjeffect}). 

It then follows from the value of the interference angle $\theta_d=141.76^{\circ}$ that the disjunction effect  can be explained as an effect of `destructive interference' created by the conceptual situations $A$ and $B$ in the combination `$A$ or $B$'. 
%M Comment: The sentence below is not clear to me, nor it is clear to me how it should be understood by the reader, hence, I suggest to explain better how it should be interpreted. Or, maybe, a reference should be added to a paper where this is explained?
It is also important to notice that uncertainty aversion can be interpreted as the effect of the overall cognitive landscape influencing the decision. Hence, effects of context play a crucial role in the disjunction effect too. 

The two-stage gamble was repeated by other scholars, finding systematic deviations from classicality. We report in {\bf Table 3} 
some of these data, together with the corresponding values from the quantum model. In addition, the quantum theoretical model above also works with the disjunction effect detected in the Hawaii problem and the prisoner's dilemma \cite{ahs2017}. 
\begin{table}
\begin{center}
\begin{tabular}{|ccccccc|}
\hline\hline
\multicolumn{7}{|c|}{\rule{0pt}{3ex}Experimental tests on the two-stage gamble version of the disjunction effect} \\
\hline\hline
\rule{0pt}{3ex} {\it Experiment}	&	$\mu(A)$	&	$\mu(B)$	&	$\mu(A \ {\rm or} \ B)$ & $\theta_{d}$	 & $|A\rangle$ & $e^{-i \theta_{d}}|B\rangle$	\\
\hline
\rule{0pt}{3ex}
{T \& S (1992)}	&	$0.69$	&	$0.58$	&	$0.37$	&	$137.26^{\circ}$	&	$(0.73,	0,	0.68)$	&	$(0.61,	0.45,	-0.66)$	\\
\hline
\rule{0pt}{3ex}
{K et al. (2001)}	&	$0.72$	&	$0.47$	&	$0.48$	&	$107.37^{\circ}$	&	$(0.85,	0,	0.53)$	&	$(0.45,	0.51,	-0.73)$	\\
\hline
\rule{0pt}{3ex}
{L \& B (2007)}	&	$0.63$	&	$0.45$	&	$0.41$	&	$106.75^{\circ}$	&	$(0.79,	0,	0.61)$	&	$(0.57,	0.36, -0.74)$	\\
\hline
\end{tabular}
\end{center}
{\bf Table 3.} {\it Quantum modeling of average data on the-stage gamble tests Tversky and Shafir (T \& S) \cite{ts1992}, K\"{u}hberger (K) et al. \cite{k2001} and Lambdin and Burdsal (L \& B)  \cite{lb2007}.} 
\end{table}

In the previous sections, 
we have briefly illustrated the modeling effectiveness of the quantum theoretical approach to human cognition we have developed in Brussels in the last decade. In the section that follows, we intend to explain where the quantum cognitive effectiveness comes from.

\section{A unifying explanatory hypothesis\label{explanation}}
We present in this section an hypothesis we have recently elaborated to explain the appearance of genuine quantum structures in cognitive phenomena. The explanatory hypothesis reveals very stable patterns of human  reasoning, enlightening at the same time some fundamental traits of its deepest nature \cite{IQSA2}.

We have extensively discussed here 
 the fundamental limitations of classical set theoretical formalisms, like those employed to formalize Boolean logic and Kolmogorovian probability, in the modeling of several cognitive phenomena. These deviations from classicality put at stake the descriptive but, to some extent also the normative, foundations of rational decision theory, according to which any source of uncertainty can be formalized probabilistically in a Kolmogorovian sense \cite{bb2012}. The Kahneman-Tversky research programme interprets the deviations above as true fallacies of human reasoning and already provides a step forward, also because it introduces, albeit implicitly and only partially, non-Kolmogorovian models of probability \cite{tk1974}. However, the explanation in terms of individual biases and judgment heuristics, though valid at an intuitive level, cannot be the definitive answer to the problem, as it does not provide any explanation in terms of human reasoning processes.

The quantum cognition research programme has the significant advantage of exploiting the modeling flexibility offered by the quantum mechanical formalism with respect to classical set theoretical formalisms. We have however seen in Section \ref{brussels} that there is no evidence that human reasoning processes can be explained in terms of microscopic quantum processes. Thus, one is naturally led to wonder how and why the quantum formalism is so effective in representing complex cognitive phenomena, like categorization, judgment, decision-making and perception?

The quantum theoretical approach to cognition illustrated here suggests a simple explanatory hypothesis on the mechanisms that underlie human reasoning. According to our explanatory hypothesis, reasoning is a specifically structured superposition of two 
processes:  `logical reasoning' and `emergent reasoning'. Logical reasoning combines conceptual entities, that is, concepts, combinations of concepts, propositions and also more complex decision entities, by applying the rules of Boolean logic, though in some cases enriched by a Kolmogorovian probabilistic semantics (see, e.g., graded membership in conceptual combinations, Section~\ref{concepts}). Emergent reasoning enables instead formation of combined conceptual entities as  newly emerging entities, in the case of concepts, new concepts, in the case of propositions, new propositions, etc., carrying new meaning, linked to the meaning of the constituent conceptual entities, but with a connection that cannot be formalized by the algebra of Boolean logic. These two mechanisms act in superposition in human thought during a reasoning process, the first one being guided by an algebra of logic, the second one following a mechanism of emergence. 

In the above perspective, human reasoning can be mathematically formalized in a Fock space, where only the first two sectors are active, and the states of conceptual entities are represented  by unit vectors of this `two-sector Fock space'. More specifically,  `sector 1 of Fock space', formed by a single Hilbert space, models `conceptual emergence', hence the combination of two concepts is represented by a superposition vector of the vectors representing the component concepts in this Hilbert space, allowing quantum interference between conceptual entities to play a role in the process of emergence. `Sector 2 of Fock space', i.e. a tensor product of two identical copies of this Hilbert space, models a conceptual combination from the combining concepts by requiring  the rules of logic for the logical connective used for the combining, i.e. conjunction or disjunction, to be satisfied in a probabilistic setting. This quantum-theoretic modeling suggested us to call `quantum conceptual thought' the process occurring in sector 1 of Fock space, `quantum logical thought' the process occurring in sector 2. 

The relative importance of emergence  or logic in a specific cognitive process is measured by the `degree of participation' of sectors 1 and 2. The abundance of evidence of deviations from classical logical reasoning in concrete human decisions, together with our results in Section \ref{concepts} on the mechanisms of concept formation identified in the conjunction and negation of natural concepts, induce to draw the conclusion that emergence constitutes the dominant dynamics of reasoning, but a dynamics of logic is systematically present too, though only at a lower level. 

Coming now to the non-classical effects detected in  cognition and analyzed in Sections \ref{concepts}, \ref{linda} and \ref{disjunction}, 
 we believe that over- and under-extension effects in typicality and membership judgments, conjunctive and disjunctive fallacies, and disjunction effects are a consequence of the dominant dynamics and their nature is emergence. This explains why the quantum formalism in Hilbert space (or sector 1 of Fock space) is so effective to model empirical data about 
these effects. It follows that paradoxes, fallacies, effects and contradictions are not `errors from a default logical reasoning' but, 
rather, natural expressions of the dominant emergent reasoning. 

On the other side, the results presented in Sections \ref{concepts}, \ref{linda} and \ref{disjunction} 
are so significant that one is naturally led to wonder whether logical aspects do play any role in the reasoning processes generating the 
different  effects. This is particularly evident in conjunctive fallacies and disjunctive effects, where all data can be represented in sector 1 of Fock space and a mechanism of conceptual emergence is sufficient to explain the observed deviations from classicality. We believe that this is due to the fact that original experiments on human judgment and decision-making, such as those by Hampton \cite{h1988a,h1988b}, Tversky-Kahneman \cite{tk1983} and Tversky-Shafir \cite{ts1992}, were explicitly designed to identify deviations from classical logical reasoning in concrete decisions, hence they on purpose `triggered' emergent aspects of human reasoning. The situation is more complex in conceptual categorization, where `pure classical data' cannot be modeled in sector 1 of Fock space alone, but they already require sector 2, as we have briefly sketched in Section \ref{concepts}. Hence, 
there are reasons to  believe that decision-making experiments can be designed where both Fock space sectors are needed to model the data, because the cognitive effect under study would trigger both aspects of human reasoning, emergent and logical,  with a similar level of participation  to the overall cognitive process.

%\bigskip
%\noindent
%{\bf Acknowledgments.} This work was supported by the Marie Sk{\l}odowska-Curie Innovative Training Network 721321 -- ``QUARTZ''.

\end{document}